\documentclass{iopjournal}

\usepackage{physics, slashed, multirow,xcolor,booktabs,bbold}
\usepackage{cite, hyperref}
\usepackage{url}

\usepackage[bitstream-charter]{mathdesign}

\hypersetup{
    colorlinks,
    linkcolor={red!50!black},
    citecolor={blue!50!black},
    urlcolor={blue!80!black}
}

\newcommand{\de}{\text{d}}

\begin{document}

\articletype{Letter} 
\begin{flushright}
\end{flushright}
\title{Revisiting the tensionless limit of pure--Ramond-Ramond AdS3/CFT2}

\author{Alberto Brollo$^{a,b}$\orcid{0009-0004-4939-516X}, Dennis le Plat$^{c}$\orcid{0000-0002-0433-7212} and Alessandro Sfondrini$^{d,e}$\footnote{Corresponding author. On leave from the University of Padova, Italy.}\orcid{0000-0001-5930-3100}}

\affil{$^a$Technical University of Munich, CIT, Department of Mathematics, Boltzmannstra{\ss}e 3, 85748 Garching, Germany}

\affil{$^b$Technical University of Munich, TUM School of Natural Sciences, Physics Department, 85748 Garching, Germany}

\affil{$^c$HUN-REN Wigner Research Centre for Physics, 1121 Budapest, Konkoly-Thege Miklós út 29-33, Hungary}

\affil{$^d$School of Mathematics, University of Birmingham,
Watson Building, Edgbaston, Birmingham B15 2TT, UK}

\affil{$^e$INFN, Sezione di Padova, via Marzolo 8, 35131 Padova, Italy}

\email{alberto.brollo@tum.de, dennis.leplat@wigner.hu, a.sfondrini@bham.ac.uk}

\begin{abstract}
We revisit the numerical solution of the mirror TBA equations for pure--Ramond-Ramond strings  on $AdS_3\times S^3\times T^4$ in the tensionless limit. Our analysis uses the recently-proposed modification of the dressing factors which account for non-trivial exchange relations of the massless modes.
At leading order in the tension, the dynamics is driven by the massless excitations associated to~$T^4$ modes and their superpartners, but it is non-relativistic and interacting unlike what happens in the symmetric-product orbifold CFT of~$T^4$. 
\end{abstract}
\vspace{\baselineskip}

\vspace{10pt}
\noindent\rule{\textwidth}{1pt}
\tableofcontents
\noindent\rule{\textwidth}{1pt}
\vspace{10pt}

\section{Introduction}
\label{sec:intro}

The duality between superstrings in $AdS_3$ backgrounds and two-dimensional Conformal Field Theories (CFT2) is a key example of holography since its inception~\cite{Maldacena:1997re,Giveon:1998ns,Kutasov:1999xu,Seiberg:1999xz,Berkovits:1999im,Maldacena:2000hw}, and it has seen a renewed interest as of late (see~\cite{Seibold:2024qkh} for a recent review and list of references). The simplest and arguably best-understood superstring background of this type is $AdS_3\times S^3\times T^4$, which preserves sixteen Killing spinors and can be supported by a mixture of Ramond--Ramond (RR) and Neveu-Schwarz--Neveu-Schwarz (NSNS) flux. Calling $G_{\mu\nu}(X)$ the metric tensor for $AdS_3\times S^3\times T^4$ where~$X^\mu$, $\mu=0,\dots, 9$ are coordinates, the classical string action takes the form
\begin{equation}
S=-\frac{T}{2}\int\limits_{-\infty}^{+\infty}\de\tau\int\limits_0^{L}\de\sigma\left[\gamma^{ij}G_{\mu\nu}(X)+\varepsilon^{ij}B_{\mu\nu}(X)\right]\partial_iX^\mu\partial_j X^\nu+\text{fermions}\,.
\end{equation}
Here $\gamma_{ij}$ is the unit-determinant worldsheet metric, $L$ is the volume of the worldsheet theory (both will be fixed by a suitable gauge choice), and $T$ is the string tension. The Kalb--Ramond field~$B_{\mu\nu}(X)$ is related to the amount of the NSNS flux. In fact, to obtain a consistent supergravity background we need to include the NSNS flux~$H_{(3)}$ and the RR flux $F_{(3)}$ which (in the conventions of~\cite{Lloyd:2014bsa}) are given in terms of the volume-forms as
\begin{equation}
    H_{(3)}=\de B=q\,\Omega\,,\qquad F_{(3)}=\sqrt{1-q^2}\,\Omega\,,\qquad
    \Omega=2\,\text{vol}(\text{AdS}_3)+2\,\text{vol}(\text{S}^3)\,,\qquad 0\leq q\leq1\,.
\end{equation}
Consistency of the string action at the quantum level requires $2\pi qT\equiv k$ to be quantised, $k\in\mathbb{N}$. Introducing a parameter $h\equiv \sqrt{1-q^2}T$ (which is related to the strength of the RR flux), we can rewrite the tensions as 
\begin{equation}
    T=\sqrt{\frac{k^2}{4\pi^2}+h^2}\,,\qquad k\in\mathbb{N}\,,\qquad h\geq0\,.
\end{equation}
This highlights how the tension is sourced from the couplings to NSNS and RR fluxes. It is worth remarking that, for any value of $k,h$ above, the action enjoys the same amount of supersymmetry --- sixteen real supercharges which close in the $\mathfrak{psu}(1,1|2)\oplus\mathfrak{psu}(1,1|2)$ superalgebra. While different values of these parameters may be related by S~duality, in perturbative string theory it is natural to treat them as different models.

In the context of holography, one would like to understand (to begin with) the spectrum of free superstrings for any value of $k,h$, and to match it to a dual super-conformal field theory (SCFT) in two dimensions with $\mathcal{N}=(4,4)$ supersymmetry. One might expect that this identification should be simplest in the limit where the tension is very small. By analogy, in the case of $AdS_5\times S^5$  superstrings, this would be the limit of free planar $\mathcal{N}=4$ supersymmetric Yang--Mills theory. In the case of $AdS_3\times S^3\times T^4$, such ``tensionless limit'' is taken to mean either
\begin{enumerate}
    \item the case where $k=1$ and $h\ll1$, related to the symmetric-product orbifold CFT of~$T^4$~\cite{Eberhardt:2019ywk}, or
    \item the case where $k=0$ and $h\ll1$, whose dual is not yet well understood,
\end{enumerate}
see also~\cite{Seibold:2024qkh} for a recent review and references. This paper is concerned with the study of the latter case (in a sense, ``the most tensionless''), where the tension is sourced only by RR fluxes and their strength is vanishingly small. Because this setup involves RR-fluxes, it is not known how to study it by means of the worldsheet-CFT techniques which prove so powerful for pure-NSNS backgrounds~\cite{Berkovits:1999im,Maldacena:2000hw}. Currently, the only known approach to this case relies on using integrability to quantise the Green--Schwarz superstring in lightcone gauge, like it was done for $AdS_5\times S^5$~\cite{Arutyunov:2009ga,Beisert:2010jr}.

The integrability approach to $AdS_3\times S^3\times T^4$ superstrings with pure-RR flux was initiated in~\cite{Babichenko:2009dk}. It relies on bootstrapping the S-matrix on the string worldsheet from symmetries~\cite{Borsato:2012ud,Borsato:2013qpa,Sfondrini:2014via}. For the most part, the S-matrix elements can be fixed by studying the linearly realised symmetries of the worldsheet model in lightcone gauge, in a suitable limit where~$L$ (which by gauge fixing is related to the $S^3$ angular momentum) is taken to be infinite and the worldsheet becomes a plane. However, some overall coefficients --- the so-called \textit{dressing factors} --- can only be fixed by imposing unitarity, parity, crossing symmetry~\cite{Janik:2006dc}, and by requiring good analytic properties. This is highly non-trivial due to the involved analytic structure of the worldsheet model, whose dispersion relation is non-relativistic,
\begin{equation}
\label{eq:dispersion}
    E_\mu(p)=\sqrt{\mu^2+4h^2\left(\sin\frac{p}{2}\right)^2}\,,\qquad \mu\in\mathbb{Z}\,.
\end{equation}
Here $\mu$ distinguishes different representations of the lightcone symmetry algebra. While the symmetry analysis of  the worldsheet S~matrix was completed in 2014~\cite{Borsato:2012ud,Borsato:2013qpa,Sfondrini:2014via}, the dressing factors required substantially more work~\cite{Frolov:2021fmj,Frolov:2025ozz} --- we will return to this point in a moment.%
\footnote{Earlier proposals for the dressing factors~\cite{Borsato:2013hoa,Borsato:2016xns} turned out to be flawed, see~\cite{Frolov:2021fmj}.}
Once the S-matrix is fully understood, one can use Bethe ansatz techniques to find the  string spectrum on a cylinder of finite spatial volume~$L$. In this worldsheet model, $L$ turns out to be integer as it is related to the R-charge of the state~\cite{Arutyunov:2009ga}.
Qualitatively, we expect the momenta of worldsheet excitations $p_j$, $j=1,\dots N$ become quantised (roughly as~$1/L$) and the energy of a multi-particle state in volume~$L$ to be
\begin{equation}
\label{eq:energy}
    \mathbf{E}|p_1,\dots p_N\rangle_L=
    E_{N,L}|p_1,\dots p_N\rangle_L\,,\qquad
    E_{N,L}=\sum_{j=1}^N E(p_j)+\text{finite-$L$ effects.}
\end{equation}
Similarly to what was done in relativistic integrable QFTs~\cite{Zamolodchikov:1989cf,Dorey:1996re}, the exact value of~$E_{N,L}$ in finite volume can be found by studying the finite-temperature \textit{thermodynamic Bethe ansatz} (TBA) equations, up to exchanging worldsheet space and time by means of a double Wick rotation:
\begin{equation}
    \tilde{\tau}= i\,\sigma\,,\qquad \tilde{\sigma}=i\,\tau\,.
\end{equation}
It is important to note that non-relativistic models such as this, cf.~\eqref{eq:dispersion}, the double Wick rotation defines a new ``mirror'' model~\cite{Arutyunov:2007tc} with rather different properties. For instance, the mirror dispersion relation is now
\begin{equation}
\label{eq:mirrordispersion}
    \tilde{E}_\mu(\tilde{p})=2\,\text{arcsinh}\left(\frac{\sqrt{\tilde{p}^2+\mu^2}}{2h}\right)\,,\qquad\mu\in\mathbb{Z}\,.
\end{equation}
This represents the energy of virtual particles which ``wrap'' the worldsheet cylinder contributing to the finite-volume effects as $\de E^{\text{wrap}}\approx\exp({-L\tilde{E}(\tilde{p})})\de\tilde{p}$.  For pure-RR $AdS_3\times S^3\times T^4$, the mirror TBA was worked out in~\cite{Frolov:2021bwp}.

The aim of this paper is to use the mirror TBA to study the spectrum of pure-RR $AdS_3\times S^3\times T^4$ superstrings in the tensionless limit, i.e.\ at leading order in~$h\ll1$.
This is important to gain information on the dual CFT in the regime where it should be weakly coupled.
From~\eqref{eq:dispersion} we see that the leading contributions at $h\ll1$ arise from $\mu=0$ modes (without a mass gap), and they are of order $\mathcal{O}(h)$. It is also easy to see that finite-volume effects $\exp({-L\tilde{E}(\tilde{p})})\de\tilde{p}$ are dominant for mirror particles with $\mu=0$ (also gapless) and that they appear at the same order $\mathcal{O}(h)$. The mirror TBA is currently the only approach which can take these finite-volume effects into account and predict the spectrum in this regime.%
\footnote{The ``Quantum Spectral Curve'' (QSC) is another system of equations for the string spectrum, akin to $QQ$-relations. In the case of $AdS_3\times S^3\times T^4$ it was put forward based on symmetry considerations, self-consistency, and an educated guess for the analytic properties of its $Q$ functions~\cite{Ekhammar:2021pys,Cavaglia:2021eqr}.
Though the study of these equations is underway~\cite{Cavaglia:2022xld,Ekhammar:2024kzp}, it is not currently known if they are equivalent to the mirror TBA and how to use them to study the $\mathcal{O}(h)$ spectrum.}

We have already considered this issue in our work~\cite{Brollo:2023pkl} (see also~\cite{Brollo:2023rgp} for further technical details of the computation).
However, recently it was suggested~\cite{Ekhammar:2024kzp,Frolov:2025ozz} that the dressing factor~\cite{Frolov:2021fmj} used in the original mirror TBA~\cite{Frolov:2021bwp} (and subsequently in our work~\cite{Brollo:2023pkl,Brollo:2023rgp}) \textit{may not be the correct one}. The issue is explained in detail in~\cite{Frolov:2025ozz}: the usual derivation of the the crossing equations assumes that particles obey standard bosonic or fermionic exchange relations (exchanging two creation operators yields a $\pm1$ prefactor) while in the case of massless excitations on $AdS_3\times S^3\times T^4$ there is reason to believe that they may obey non-standard ``semionic'' relations (exchanging two creation operators yields a $\pm i$ prefactor). This is not uncommon in two-dimensional (integrable) QFTs. For $AdS_3\times S^3\times T^4$, the semionic exchange relations mean that a different and simpler (or more precisely, ``more minimal'') solution of the crossing equation is allowed.

Hence, in a nutshell the aim of this letter is to repeat the $h\ll1$ analysis of~\cite{Brollo:2023pkl,Brollo:2023rgp} when the mirror TBA of~\cite{Frolov:2021bwp} is amended with the massless dressing factors conjectured in~\cite{Ekhammar:2024kzp,Frolov:2025ozz}. This can be done by taking the mixed-flux mirror TBA equations recently proposed in~\cite{Frolov:2025tda} and setting $k=0$, $h\ll1$. It so happens that in this case the convergence of the resulting TBA is more subtle than in~\cite{Brollo:2023pkl,Brollo:2023rgp}, and requires some additional care. We will start by revisiting the tensionless limit of the mirror TBA equations in section~\ref{sec:tensionless}, and discuss the numerical evaluation in section~\ref{sec:numerics}. Some details on the convergence of the numerical algorithm are presented in appendix~\ref{app:numerics}, and the code is available in Zenodo~\cite{Zenodo}. Finally, we comment on our results and their significance in section~\ref{sec:conclusions}.

\section{Revisiting the tensionless limit}
\label{sec:tensionless}

Let us briefly review the key points from the derivation of the simplified TBA equations for $AdS_3\times S^3\times T^4$, referring the reader for~\cite{Brollo:2023rgp} for the definitions of the various quantities relevant for the TBA equations.
As a starting point we take the mirror TBA equations of~\cite{Frolov:2025tda} with $k=0$ and $h\ll1$.  They involve an infinite tower of massive modes, with Y-functions~$Y_Q$ ($Q=1,2,3,\dots$), $N_0$ massless modes with Y-function $Y_0$, and two sets of auxiliary roots denoted by~$Y_\pm$. Currently, it is an open question whether $N_0=1$ or $N_0=2$;%
\footnote{Na\"ively, one would expect two identical sets of massless modes with the same equations, $N_0=2$, as a result of the $\mathfrak{su}(2)_\circ$ structure of the S~matrix~\cite{Borsato:2014exa}. However, matching of the ground-state TBA equation with twisted boundary conditions~\cite{Frolov:2023wji} suggests that it should be~$N_0=1$.}
 we will see below that our weak-tension analysis indicates that, in fact, it should be $N_0=1$.
By using the contour-deformation approach of~\cite{Dorey:1996re}, the ground-state TBA equations can be promoted to excited-state ones. We are interested in exciting massless modes by picking up singularities in the TBA equations corresponding to massless particles in the string kinematics.
This adds to the TBA equations source terms involving the S~matrix in the string--mirror kinematics and simultaneously imposing the \emph{exact Bethe equations} determines the value of the Bethe roots; these equations involve the string--string S~matrix between massless modes.
Because this integrable model is a worldsheet QFT, the particles must obey the level-matching condition
\begin{equation}
\label{eq:lvlmatch}
    \sum_{j=1}^{N}p_j=0\quad\text{mod}~2\pi\,.
\end{equation}
Due to this constraint, it is more interesting to consider $N$-particle states with $N\geq2$. 
In this paper, we excite $M$~pairs of massless modes for a total of $N=2M$ particles.
These correspond to excitation of the fermionic superpartners of the $T^4$ bosons of the $AdS_3\times S^3\times T^4$ Green--Schwarz superstring.
We take their momenta to come in pairs $\{p_j,p_{j-1}\equiv-p_j\}$, with $j=2,4,\dots, 2M$, ensuring that~\eqref{eq:lvlmatch} is trivially satisfied. Moreover, in this way Y-functions appearing in the TBA equation will always be even, which simplifies the numerics.

\subsection{Decoupling of massive $Y_Q$ functions}

An essential simplification arises when considering the small-tension limit $h \ll1 $. The massive mirror energy~\eqref{eq:mirrordispersion} for a $Q$-particle bound state ($Q=|\mu|$) behaves as $\tilde{E} \sim \ln({Q}/{h})$, so the massive $Y$-functions are suppressed as $Y_Q \sim h^{2L}$, with $L$ being the length of the state.
This is not the case for the massless Y-functions $Y_0$ and for the auxiliary ones~$Y_\pm$, which are of order~$\mathscr{O}(h^0)$.
Considering the leading order at~$h\ll1$, we see that TBA kernels relating massive modes and auxiliary ones are regular in the limit, so that indeed massive modes decouple from the auxiliary TBA equations, which read
\begin{equation}
\begin{aligned}
\label{eq:Ypm_renorm}
    \ln{Y_-} =& + \ln{(1+Y_0)^{N_0}} \check{\star} K^{0y} 
    - \sum_{j=1}^{2M} \ln S^{0_*y}(u_j,\,\cdot\,)\,, \\
    \ln{Y_+} =& -  \ln{(1+Y_0)^{N_0}} \check{\star} K^{0y}
    + \sum_{j=1}^{2M} \ln S^{0_*y}(u_j,\,\cdot\,)\,,
\end{aligned}
\end{equation}
where the various Kernels and convolutions can be found in~\cite{Brollo:2023rgp}.
The two equations are opposite to each other, so that we may set
\begin{equation}
    Y\equiv Y_+ = \frac{1}{Y_-}\,.
\end{equation}
Using this simplification and observing that at $h\ll1$ the massive~$Y_Q$ decouple from the $Y_0$  equation too, we have
\begin{equation}
    -\ln Y_0 = L \tilde{E}_0 -  \ln{(1+Y_0)^{N_0}} \check{\star} K^{00}_{\text{ren}}- \ln{\left(1-Y\right)^4} \hat{\star} K^{y0} +  \sum_{j=1}^{2M} \ln S^{0_*0}_{\text{ren}}(u_j,\,\cdot\,) \,, 
\label{eq:Y0_renorm}
\end{equation}
where the renormalized kernel $K^{00}_{\text{ren}}$ is given by~\cite{Brollo:2023rgp}
\begin{equation} \label{eq:RenormK00}
    K^{00}_{\text{ren}} = K^{00} + 2 K^{0y} \hat{\star} K^{y0} \,, \qquad \text{with} \qquad K^{QP}(u,v)= \frac{1}{2\pi i} \frac{\de}{\de u} \ln S^{QP}(u,v).
\end{equation}
Continuing the massless TBA equation \eqref{eq:Y0_renorm} to the string region, we obtain the exact Bethe equations labeled by $k=1,\dots,2 M$,
\begin{equation}
\label{eq:SimplifiedExactBetheEqs}
    2\pi i\,\nu_k= -i L p(u_k) -  \left(\ln{(1+Y_0)^{N_0}} \check{\star} K^{00_*}_{\text{ren}}\right)(u_k)- \left(\ln{\left(1-Y\right)^4} \hat{\star} K^{y0_*}\right)(u_k)
    + \sum\limits_{\substack{j=1 \\ j\neq k}}^{2M} \ln S^{0_*0_*}_{\text{ren}}(u_j,u_k)\,,
\end{equation}
 with the asterisks indicating the continuation to the string kinematics and where we used that, for coincident rapidities, $S^{0_*0_*}_{\text{ren}}(u,u)=-1$. Reassuringly, the $h\ll1$ limit and the analytic continuation turn out to be commuting operations.
Based on these equations, $Y_0$, $Y$ and $p_j$ ($j=1,\dots, 2M$) are all of order $\mathcal{O}(h^{0})$ as $h\ll1$.
Finally, at $h\ll1$ the energy is given by
\begin{equation}
    E_{M,L}=-\frac{1}{2\pi}\int\limits_{-\infty}^{+\infty}\de\tilde{p}\, \ln(1+Y_0(\tilde{p}))^{N_0}+\sum_{j=1}^{2M}E(p_j)\,,
\end{equation}
where again the contribution of $Y_Q$ functions is subleading and omitted. As we will see in the next subsection, both of these terms are of order~$\mathcal{O}(h^1)$.

\subsection{Difference-form small-tension TBA equations}
Because only massless and auxiliary Y-functions play a role in our analysis, it is convenient to parametrise them by the $\gamma$-rapidity variable introduced in~\cite{Fontanella:2019baq,Frolov:2021fmj}.%
\footnote{The same parametrisation also appeared in the study of the $AdS_5\times S^5$ dressing factors~\cite{Beisert:2006ib}.}
In terms of this, the TBA equations become of difference form --- though they are still non-relativistic.

Let us now consider the massless scattering phase proposed for the mixed-flux \cite{Frolov:2025tda} given by
\begin{equation}
    S^{00}(u_j,u_k) = - \Phi_{\text{SG}} (\gamma_{j}-\gamma_k)^2\, (\Sigma_{\text{BES}}^{0 0}(\gamma_j,\gamma_k))^{-2} \,,\qquad  K^{00} = 2 K_{\text{SG}} - 2 K_{\text{BES}}^{0 0}\,,
\label{def:S00}
\end{equation}
in term of the Beisert--Eden--Staudacher (BES)~\cite{Beisert:2006ez} kernel in the massless mirror kinematics~\cite{Frolov:2021fmj,Frolov:2025tda}.
The convolution 
\begin{equation}
    (K^{0y} \hat{\star} K^{y0})(u,u')  = \left( \frac{\de u}{\de  \gamma} \right)^{-1} \left[  - K_{\text{SG}} (\gamma- \gamma') + \frac{1}{4} \delta (\gamma) \right] \,,
\end{equation}
was worked out in \cite{Brollo:2023rgp}. Therefore, the renormalized kernel \eqref{eq:RenormK00} reads $K^{00}_{\text{ren}}= - 2 K_{\text{BES}}^{0 0}$.%
\footnote{The contribution of the $\delta$-function can be neglected because, due to the driving term $L\tilde{E}_0(\gamma)$, the function $Y_0(\gamma)$ vanishes at~$\gamma=0$, which is the support of~$\delta(\gamma)$ in the TBA equations. One might worry that the $\delta(\gamma)$ kernel may nonetheless be reflected in a discontinuity in the driving terms; note however that an expression of the type $\tfrac{1}{2}\sum_k i\pi\text{sgn}(\gamma_k)$ would cancel for the states which we are considering here.}
Note that \textit{this is different from the massless kernel in~\cite{Brollo:2023pkl,Brollo:2023rgp}}. In that case, the kernel $K^{00}_{\text{ren}}$ also involved the Cauchy kernel 
\begin{equation}
    s(\gamma)= \frac{1}{2\pi \cosh{\gamma}}\,,\qquad
    s(\gamma)=\frac{1}{2\pi i}\frac{\de}{\de\gamma}\ln S(\gamma)\,,\qquad
    S(\gamma)=-i \tanh(\frac{\gamma}{2}-\frac{i \pi}{4} )\,.
\end{equation}
In \cite{Brollo:2023pkl,Brollo:2023rgp} it was shown that the convolution with the BES kernel $\ln (1+Y_0) * K_\text{BES}^{00}$ does not contribute to the equation for massless particles at $\mathcal{O}(h^0)$. The same is true for the driving terms involving the BES kernel in the string--mirror and string--string kinematics.
Thus, these terms drop from the simplified TBA equations, which read 
\begin{equation}
\label{eq:SimplifiedTBA_excited}
\begin{aligned}
\ln Y_0(\gamma) &= -L \tilde{E}_0(\gamma) + \left(\ln(1-Y)^4 * s\right)(\gamma)\,, \\
\ln Y(\gamma) &= \left(\ln (1+Y_0)^{N_0} * s\right)(\gamma) + \sum_{j=1}^{M} \ln \left(S_*(\gamma_{2j} - \gamma) S_*(-\gamma_{2j} - \gamma)\right)\,,
\end{aligned}
\end{equation}
with the exact Bethe equations and the energy given by
\begin{equation}
\label{eq:simplifiedExactBetheEnergy}
    2\pi i\nu_k= -i L p(\gamma_k) - \left(\ln{\left(1-Y\right)^4}  * s_*\right)(\gamma_k)\,,\qquad
E_{M,L}=-\int\limits_{-\infty}^{+\infty}\frac{\de\gamma}{2\pi}\,\frac{\de\tilde{p}}{\de\gamma} \ln(1+Y_0(\gamma))^{N_0}+\sum_{j=1}^{2M}E(\gamma_j)\,,
\end{equation}
where 
$f_*(\gamma)\equiv f(\gamma+\frac{i \pi}{2})$ indicates the analytic continuation to the string--mirror region, so that
\begin{equation}
s_*(\gamma)=\,\frac{1}{2\pi i \sinh(\gamma)},
\qquad S_*(\gamma)=-i \tanh(\frac{\gamma}{2})\,,
\end{equation}
similarly to a relativistic model. The energy and momentum however are non-relativistic, namely,
\begin{equation}
    \tilde{E}_0(\gamma)=-\ln\left(\frac{1-e^{\gamma}}{1+e^{\gamma}}\right)^2\,,\qquad
    p(\gamma)=-i\ln\left(\frac{e^{\gamma}-i}{e^{\gamma}+i}\right)^2,\qquad
    E(\gamma)=\frac{2h}{\cosh\gamma}\,,\qquad
    \frac{\de\tilde{p}}{\de\gamma}=\frac{2h\cosh\gamma}{(\sinh\gamma)^2}\,.
\end{equation}

\begin{figure}[t]
    \centering
    \includegraphics[width=\linewidth]{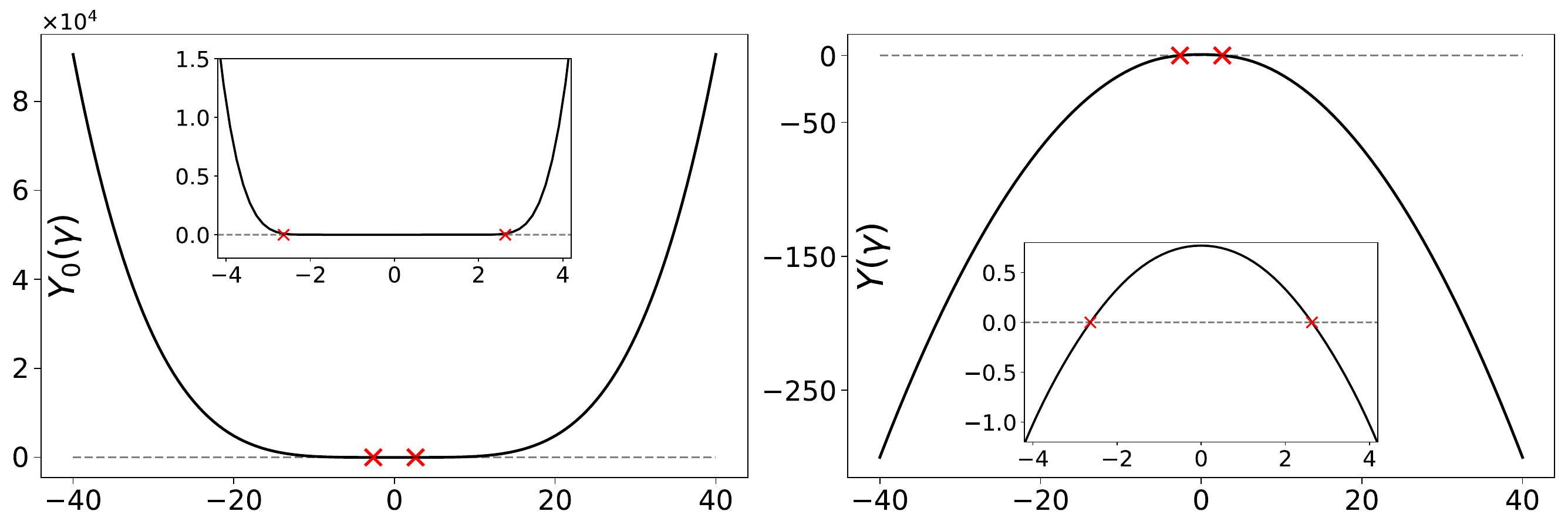}
    \caption{\textbf{Y-functions with divergent asymptotics.} A new feature of the revised TBA equations is that, for odd $M$, the Y-functions exhibit the asymptotic behaviour~\eqref{eq:divergentY}. We plot the numerical solution for $Y_0(\gamma)$ and $Y(\gamma)$ for $M=1$. The red crosses indicate the positions of the Bethe roots $\gamma_1$ and $\gamma_2=-\gamma_1$, which are zeros of the auxiliary Y-function $Y(\gamma)$. The insets provide a zoomed-in view around~$\gamma=0$.}
    \label{fig:asymptotics}
\end{figure}

\subsection{Asymptotic solutions}

In preparation for the numerical solution of~\eqref{eq:SimplifiedTBA_excited}, let us work out the solution at $|\gamma|\gg1$ where the contribution of the driving terms is less important. In fact, note that
\begin{equation}
e^{-L \tilde{E}_0(\gamma)}=1+\mathcal{O}(e^{-|\gamma|})\,,\qquad
    \prod_{j=1}^M S_*(+\gamma_j - \gamma)S_*(-\gamma_j - \gamma)=(-1)^M+\mathcal{O}(e^{-|\gamma|})\,,
\end{equation}
for $\gamma_j$ fixed and finite, and $|\gamma|\gg1$.
We can now look for asymptotic solutions (valid at large~$|\gamma|$), distinguishing the cases $N_0=1$ and $N_0=2$ and keeping track of whether $M$ is even or odd.

Na\"ively, either value $N_0=1,2$ may admit constant solutions, but it should be checked that indeed these are real and stable when used as a starting point for the numeric evaluation of the equations~\eqref{eq:SimplifiedTBA_excited}. This should hold for any value of~$M$, i.e.\ for any number of excited states. To study asymptotic solutions which grow as $|\gamma|\to\infty$, it is convenient to rewrite the TBA equations~\eqref{eq:SimplifiedExactBetheEqs} as a Y-system
\begin{equation}
\label{eq:Ysys}
    Y_0\left(\gamma+ \frac{i \pi}{2}\right)\, Y_0\left(\gamma- \frac{i \pi}{2}\right) = \left(1-Y(\gamma)\right)^4\,, \qquad
    Y\left(\gamma+ \frac{i \pi}{2}\right)\, Y\left(\gamma- \frac{i \pi}{2}\right) = \left(1+Y_0(\gamma)\right)^{N_0}\,.
\end{equation}

\paragraph{Case $N_0=1$:}
For $N_0=1$ and $M$ odd, we find no constant solutions of~\eqref{eq:SimplifiedTBA_excited} at $|\gamma|\gg1$. Instead, we find the solution
\begin{equation}
\label{eq:divergentY}
    Y_0(\gamma) =\frac{4}{\pi^4}\left(|\gamma|- c\right)^4+ \mathcal{O}(\gamma^{-1})\,,\qquad
    Y(\gamma) = \frac{1}{2}- \frac{2}{\pi^2} \left(|\gamma|-c\right)^2+ \mathcal{O}(\gamma^{-1})\,,
\end{equation}
where the sub-leading coefficient $c$ is state-dependent and we assumed that the $Y$-functions are even under $\gamma\to-\gamma$ (which is going to be the case for our symmetric choice of Bethe roots). This solution is stable when taken as a starting point of the full numerical solution for any $\gamma\in\mathbb{R}$. An example of such a numerical solution is plotted in  figure~\ref{fig:asymptotics}.  Because $Y_0(\gamma)$ and $Y(\gamma)$ diverge as $\gamma\to\pm\infty$, the  numerical evaluation of the equations requires more care, as we will see.
For $N_0=1$ and $M$ even, instead, we have a candidate constant solution $Y_0=0$, $Y=1$, which turns out to be numerically stable. Instead, no solution with divergent $Y(\gamma)$ is numerically stable in this case.

\paragraph{Case $N_0=2$:}
In the case $N_0=2$, and $M$ odd, the constant solutions to~\eqref{eq:SimplifiedTBA_excited} are complex, and we rule them out. Furthermore, it is easy to see from~\eqref{eq:Ysys} that there are no solutions where $Y_0(\gamma)$ and $Y(\gamma)$ are given by a divergent function for $|\gamma|\gg1$ that is regular in the physical strip $|\text{Im}[\gamma]|\leq i\pi/2$ (such as a polynomial).
This is a further indication that the physical value of $N_0$ is $N_0=1$, as already suggested by an analysis of the twisted vacuum~\cite{Frolov:2023wji}. 

\section{Numerical treatment and results}
\label{sec:numerics}
Since the case of $N_0=2$ appears to be inconsistent, we restrict ourselves to $N_0=1$. 
The coupled system of TBA and exact Bethe equations is solved using standard iterative techniques like those described in~\cite{Brollo:2023rgp}, to which we refer the reader. Here we will highlight the new features with respect to that treatment, which have to do with the asymptotic behaviour of eq.~\eqref{eq:divergentY}.

\subsection{Numerical algorithm}
Like in~\cite{Brollo:2023rgp}, we truncate the rapidity domain $|\gamma|\leq\Lambda$ and discretize it into $\Gamma$ points. We compute convolutions with the fast Fourier transform (FFT) algorithm, and solve the equations  iteratively until convergence. However, as discussed above, for odd values of $M$ the equations display now a non-trivial asymptotics~\eqref{eq:divergentY} with a polynomial growth at infinity. As a consequence, the rapidity space cannot longer be safely discretized, and the FFT would introduce non-negligible errors in the proximity of the cutoff $|\gamma|\sim\Lambda$. On top of that, the large-$\gamma$ behaviour depends on the non-universal parameter~$c$ which we cannot fix by the asymptotic analysis. Since the driving terms of the TBA equations are exponentially localized within a finite region, we implemented a numerical scheme in which the polynomial asymptotics are factored out. As numerical errors are expected to accumulate near the cutoff, at each iterative step we extract the non-universal parameter $c$ via a fit. This is done in a region where the driving terms no longer contribute and that is well-separated from the cutoff~$\Lambda$. Although this approach leads to a significant slowdown of the algorithm, it yields high numerical accuracy, with energies and Bethe roots determined with a precision better than $10^{-9}$, see also appendix~\ref{app:numerics} for further details. Working Python codes and raw data are collected in the Zenodo folder~\cite{Zenodo}.

\begin{figure}[t]
    \centering
    \includegraphics[width=\linewidth]{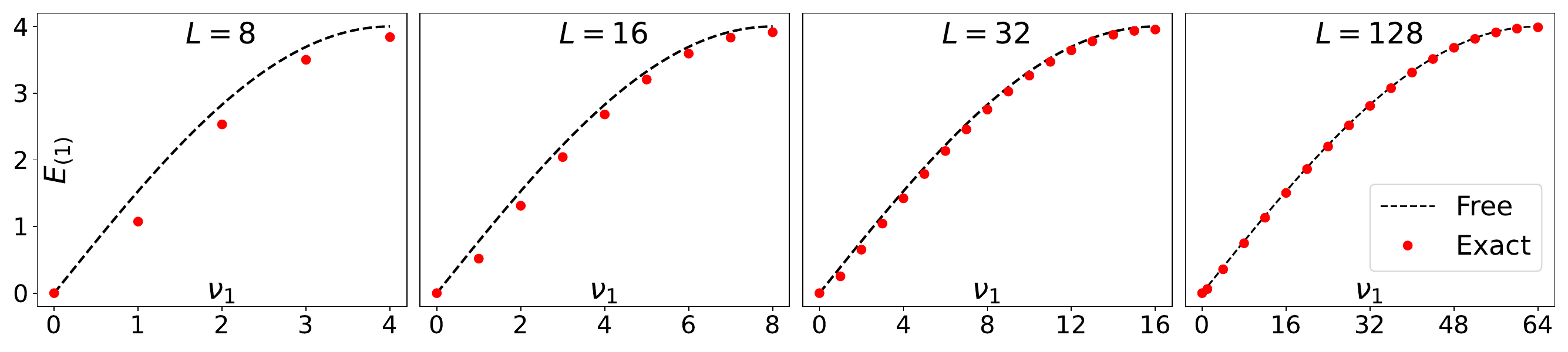}
    \caption{\textbf{Energies of two-particle states.} We consider two particle states ($M=1$) with $\nu_1=-\nu_2$ and plot their energy as function of the mode numbers $\nu_1$ for various values of the volume~$L$. The dashed line is the result obtained by discarding the finite-volume effects, cf.~\eqref{eq:free}. Note that the energy for $\nu_1=\nu_2=0$ is protected in accordance with \cite{Baggio:2017kza}.}
    \label{fig:energiesM1}
\end{figure}

\begin{figure}[t]
  \centering
\begin{minipage}{0.48\linewidth}
    \includegraphics[width=0.95\linewidth]{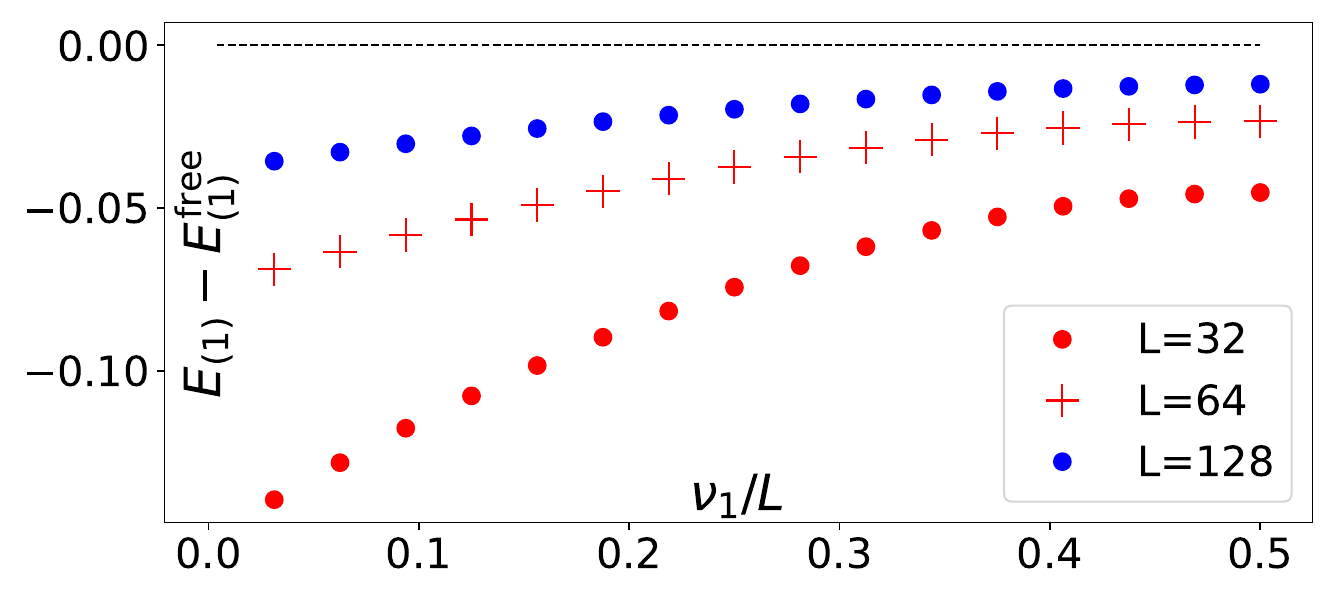}
    \caption{\textbf{Finite-size corrections as~$L$ varies.} We plot the deviation of the energy of two-particle states ($M=1$) from the free approximation.}
    \label{fig:wrapping}
\end{minipage}
\begin{minipage}{0.48\linewidth}
  \centering
    \includegraphics[width=0.95\linewidth]{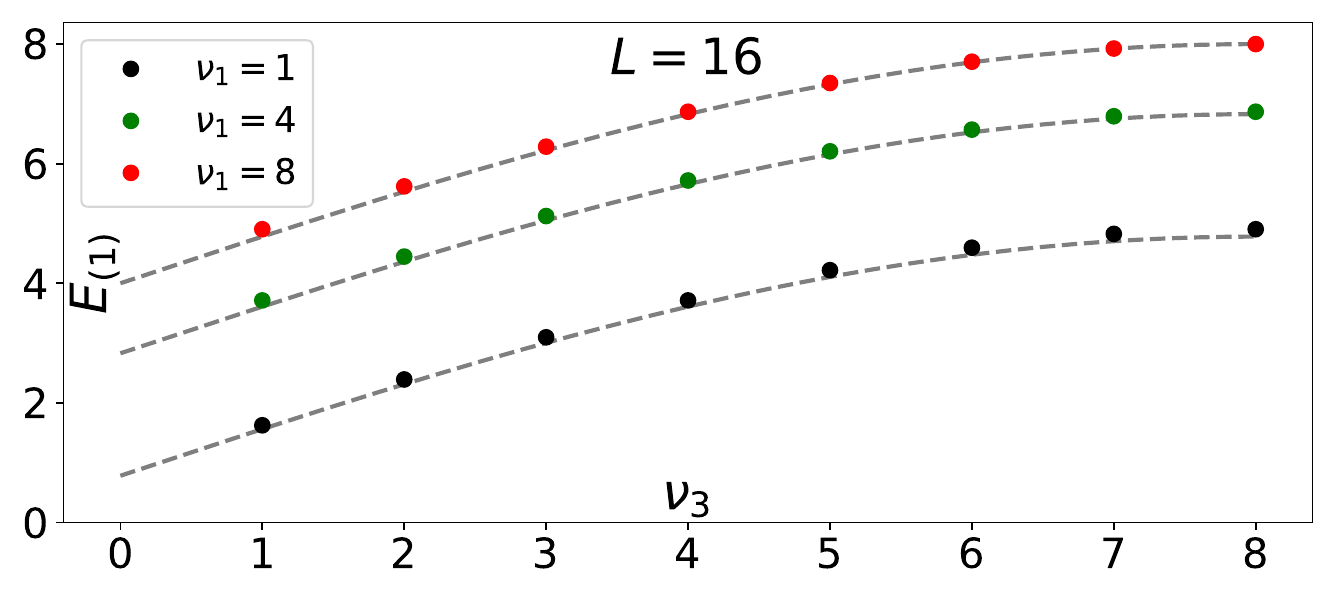}
    \caption{\textbf{Energies of four-particle states.} We plot some energies of four-particle states ($M=2$) with $\nu_1=-\nu_2$ and $\nu_3=-\nu_4$; each trajectory has fixed~$\nu_1$.}
    \label{fig:energiesM2}
\end{minipage}
\end{figure}

\subsection{Numerical results}

In this section we present the result of our numerical evaluation. As discussed around~\eqref{eq:energy}, the mirror TBA accounts for the ``asymptotic'' result and for all wrapping corrections. Expanding the complete energy at $h\ll1$ we find
\begin{equation}
    E_{2M,L}=\sum_{j=1}^{2M} E(p_j)+\text{finite-$L$ effects}=h\, E_{(1)}^{2M,L}+\mathcal{O}(h^2)\,.
\end{equation}
The equations~\eqref{eq:SimplifiedExactBetheEqs} compute precisely $E_{(1)}^{2M,L}$. Its value depends on the length $L$ (R-charge) of the state, on the number of excitations $2M$, and on their mode numbers. In practice, here we consider only two cases: $M=1$, with mode numbers $\nu_1=-\nu_2$; and $M=2$, with mode numbers $\nu_1=-\nu_2$ and $\nu_3=-\nu_4$. We will compare these results with those which we would have obtained when neglecting the wrapping corrections. To do this, let us simply drop the integrals and convolutions from~\eqref{eq:simplifiedExactBetheEnergy}, finding
\begin{equation}
\label{eq:free}
    p_{\text{free}}(\gamma_k)=-\frac{2\pi\nu_k}{L} \,, \quad k=1,\dots,2M\,,\qquad
    E_{M,L}^{\text{free}}=\sum_{j=1}^{2M}E(\gamma_j) =2h\sum_{j=1}^{2M}\left|\sin\frac{\nu_k\pi}{L} \right|\,.
\end{equation}

\begin{figure}[t]
    \centering
    \includegraphics[width=\linewidth]{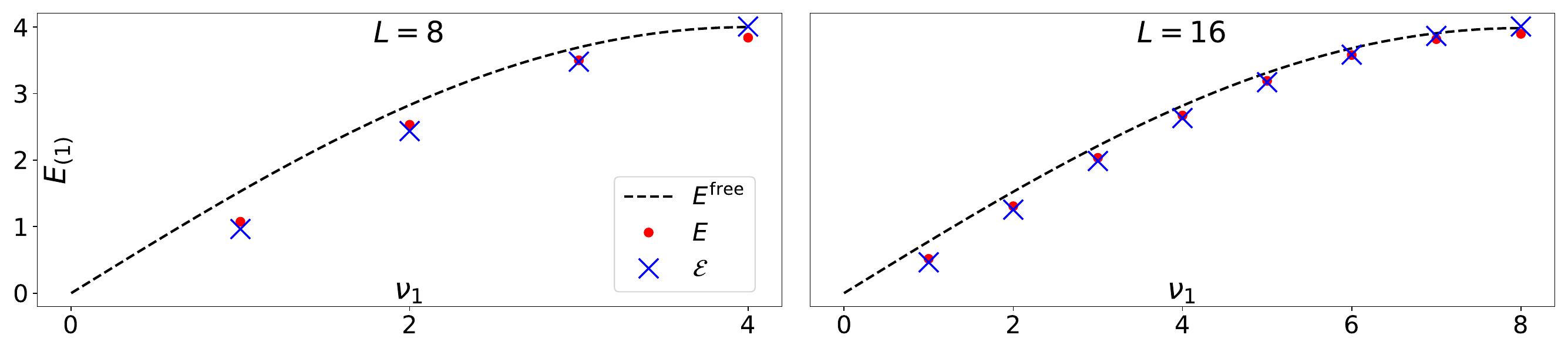}
    \includegraphics[width=\linewidth]{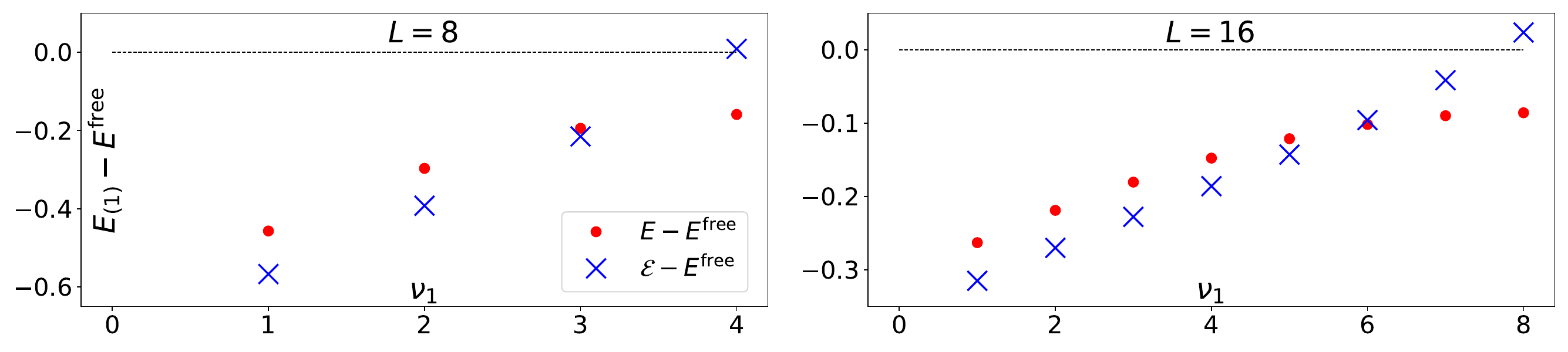}
    \caption{\textbf{Comparison with the previous results.} We compare with the results of~\cite{Brollo:2023pkl,Brollo:2023rgp}, which used a different dressing factor. In the top two plots we compare the energies, while in the bottom two we compare their deviation from the free result~\eqref{eq:free}. We indicate the results of~\cite{Brollo:2023pkl,Brollo:2023rgp} with $\mathcal{E}$ (blue crosses) and the current ones with~$E$ (red dots).}
    \label{fig:comparison}
\end{figure}

Recall that we are only discussing the case $N_0=1$.
Postponing to appendix~\ref{app:numerics} a more detailed discussion of the convergence of our algorithm, we note that our results have a numerical precision of about~$10^{-9}$. This can be achieved easily for states with $M=2$ but requires more computer time for states with~$M=1$ due to the growing asymptotics~\eqref{eq:divergentY}. 
For $M=1$ states, we plot some of the results in Figure~\ref{fig:energiesM1} for various values of~$L$. Recall that~$L$, which is the volume of the model in uniform lightcone gauge~\cite{Arutyunov:2009ga}, must be an integer because it is the R-charge of the multiplet. Overall, the exact energies deviate a little from the ones which we would encounter in a free model~\eqref{eq:free}. The deviation is not \textit{parametrically small} because we have no expansion parameter left. It is however interesting to check how large the deviation is when considering states in larger and larger volume~$L$. This quantifies the amount of wrapping corrections as a function of the volume. From figure~\ref{fig:wrapping} we see that they decrease as $1/L$, which is the expected behaviour for models with gapless excitations.
In figure~\ref{fig:energiesM2} we plot the energies for states with $M=2$, that is with four excitations having $\nu_1=-\nu_2$ and $\nu_3=-\nu_4$. The deviation from the free result is again relatively small, though it is interesting to note that in this case the exact energy is higher, rather than lower, than the one of the free model.
We compare our results with those obtained in~\cite{Brollo:2023pkl,Brollo:2023rgp}. As we can see in figure~\ref{fig:comparison}, the difference is small but noticeable. In particular, in the case considered here, the energies of two-particle states seem to always be lower than the ones expected in a free theory, whereas in~\cite{Brollo:2023pkl, Brollo:2023rgp} that depended on the mode number.

Finally, let us remark that while our approach is necessarily numerical, there are a few special configurations for which analytic statements are possible. In the case of~$\nu_1=\nu_2=0$, we find a singular solution with momenta $p_1=p_2=0$, which makes the corresponding rapidities $\gamma_1,\gamma_2$ diverge. In this case one can introduce a regulator $\varepsilon>0$ and set $p_j=\pm\varepsilon$ so that $\gamma_j=\pm\ln(\varepsilon/4)+\mathcal{O}(\varepsilon^2)$. In this way, one can check that the TBA equations~\eqref{eq:SimplifiedExactBetheEqs} can be solved by setting $Y=1+\mathcal{O}(\varepsilon)$ and $Y_{0}=\varepsilon^2+\mathcal{O}(\varepsilon^0)$. Substituting this in the energy equation~\eqref{eq:simplifiedExactBetheEnergy} and taking the limit $\varepsilon\to0$, it is easy to see that the energy of this configuration is precisely zero. This is expected because these states are protected and they correspond to half-BPS representations~\cite{Baggio:2017kza}.
Another peculiar case is that of $\pm\nu_j=L/2$. In this case, for a two-particle state, we have $p_k=\pm\pi$, corresponding to $\gamma_{j}=0^\pm$. 
In this case, we find that the convolution in the exact Bethe equations~\eqref{eq:simplifiedExactBetheEnergy} vanishes identically, because $\ln(1-Y)$ is even under $\gamma\to-\gamma$ while $s_*(\gamma-\gamma_j)$ is odd if $\gamma_j=0$. Hence, the Bethe roots are ``frozen'' at their free value~$\gamma_{j}=0^\pm$. Nonetheless, the energy is corrected by finite-volume effects, as can also be seen from figure~\ref{fig:energiesM1} and~\ref{fig:wrapping}.

\section{Conclusions and outlook}
\label{sec:conclusions}
We revisited the numerical solution of the mirror TBA equations at weak tension for pure-RR $AdS_3\times S^3\times T^4$ strings (with $k=0$ and $h\ll1$). In contrast to our previous work~\cite{Brollo:2023pkl,Brollo:2023rgp}, we  modified the massless dressing factor as recently proposed in~\cite{Ekhammar:2024kzp,Frolov:2025ozz}. From this study, we can draw two main conclusions:
\begin{enumerate}
    \item The resulting spectrum is different from that of~\cite{Brollo:2023pkl,Brollo:2023rgp} but qualitatively similar. It still relates to a supersymmetric theory of $T^4$ excitations only, interacting and non-relativistic. In particular, the spectrum is substantially different from that of the symmetric-product orbifold CFT of~$T^4$, which appears for $k=1$ and $h=0$~\cite{Eberhardt:2019ywk}.
    \item There was some confusion in the literature on whether the two massless representations of the symmetry algebra appearing in the S~matrix should be treated as one in the mirror TBA. As discussed in~\cite{Frolov:2021bwp}, when identifying them with separate Y-functions, those would obey the very same equations. As a result, treating the Y-functions as distinct amounts to introducing the factor $N_0=2$ in the mirror TBA equations, cf.\ e.g.~\eqref{eq:Ypm_renorm}. If instead we treated the two Y-functions as one, we would set $N_0=1$. It was later observed that to reproduce the twisted ground-state energy from the TBA~\cite{Frolov:2023wji} it is necessary to set $N_0=1$. Our analysis shows that, with the new dressing factors, the mirror TBA equations have solutions for $N_0=1$ but not for $N_0=2$, which allows us to rule out this possibility.
\end{enumerate}
Our work paves the way for several future explorations. Technically, it provides a testing ground to check whether the Quantum Spectral Curve (QSC) matches with the mirror TBA. There has been some recent progress in understanding how to describe massless excitations within the QSC~\cite{Ekhammar:2024kzp}, but not yet to the point of extracting predictions for the states considered here.
More physically, our work allows insights of the hitherto unknown dual of $k=0$ and $h\ll1$ superstrings. It should be a model of interacting torus excitations and their supersymmetric partners, with an underlying relativistic dynamic. A proposal for the weakly-coupled dual was put forward in~\cite{OhlssonSax:2014jtq}. Conceptually, the derivation of~\cite{OhlssonSax:2014jtq} considers the D1--D5 brane system but overlooks the torus structure. While there may be hope that this can be justified at weak tension (and for the sector of the spectrum which does not include wrapping or momentum modes on~$T^4$), this needs to be better understood.%
\footnote{We thank Edward Witten for discussions on this point.}
Regardless, this letter provides the means for quantitative tests of any proposed dual --- we stress again that the Python code and our raw data is publicly available in the Zenodo folder~\cite{Zenodo} ---  including the one discussed in~\cite{OhlssonSax:2014jtq}. Unfortunately (but unsurprisingly) in that model the structure of interactions in the massless sector is very nonlocal (in stark contrast with the nearest-neighbour interactions of weakly-coupled planar $\mathcal{N}=4$ supersymmetric Yang--Mills theory~\cite{Minahan:2002ve}) which substantially complicates the perturbative evaluation.

It would be very interesting to repeat this study for mixed-flux backgrounds, i.e.\ for $k\in\mathbb{N}$ fixed and $h\ll1$. In that case, the worldsheet dispersion relation is~\cite{Hoare:2013lja,Lloyd:2014bsa}
\begin{equation}
\label{eq:mixeddispersion}
    E(p)=\sqrt{\left(\mu+\frac{k}{2\pi}p\right)^2+4h^2\left(\sin\frac{p}{2}\right)^2}\,,
\end{equation}
and dressing factors~\cite{Frolov:2024pkz,Frolov:2025uwz} and mirror TBA equations~\cite{Frolov:2025tda} have been recently worked out. First of all, this would be an important check of the mirror TBA equations in the strict $h\to0$ limit, where the result can be found by worldsheet CFT techniques~\cite{Maldacena:2000hw}. More excitingly, at the next orders in $h\ll1$ this would provide a new and powerful method to incorporate RR flux in the spectrum, alternative to string field theory~\cite{Cho:2018nfn}. In the special case of $k=1$, where the dual at $h=0$ is known to be the symmetric-product orbifold CFT of~$T^4$~\cite{Eberhardt:2019ywk}, the TBA results could be matched with conformal perturbation theory --- see~\cite{Fabri:2025rok} for a detailed discussion of the open challenges in identifying integrability in the symmetric-product orbifold.
This is certainly exciting but it should be noted that, unlike the RR case where only $\mu=0$ contributed at $\mathcal{O}(h)$, in~\eqref{eq:mixeddispersion} an excitation with momentum $p=-2\pi\mu/k+\mathcal{O}(h)$ yields an $\mathcal{O}(h)$ contribution to the energy for any~$\mu$ (see also~\cite{Frolov:2023lwd} for more details on this limit). The analysis of the mirror dispersion relation is more involved (the model is non-unitary~\cite{Baglioni:2023zsf}) which highlights how the study of the $h\ll1$ for $k\neq0$ is likely to be technically more challenging than the study which we presented here.

Finally, we note that there has been recent progress in the study of the $AdS_3\times S^3\times S^3\times S^1$ spectrum with RR flux, with proposals for the Quantum Spectral Curve for pure-RR models~\cite{Cavaglia:2025icd,Chernikov:2025jko} (in the special case where the two spheres have the same radius), and for the massive dressing factors and string hypothesis for the mixed-flux model~\cite{Frolov:2025syj} (for general radii). The latter forms the basis of the mirror TBA. We believe that, in either approach, more study is needed: the QSC seems to be at odd with requiring both crossing symmetry and unitarity (as one usually does) while the study of the dressing factors is not yet complete, and the S-matrix fusion exhibits some unusual and awkward features. Nonetheless, we expect that in the pure-RR case the argument for the decoupling of the massive Y-functions which we presented above would hold in a similar way, allowing for a relatively easy numerical investigation of the $\mathcal{O}(h)$ spectrum.
We hope to return to some of these questions in the near future.

\section*{Acknowledgements}
We thank Sergey Frolov, Davide Polvara, and Ryo Suzuki for  fruitful collaboration on closely related topics and useful discussion. We also thank Patrick Dorey and Roberto Tateo for useful related discussions.

\paragraph{Funding information}
DlP is grateful for the funding through a Feodor Lynen Fellowship by the Alexander von Humboldt Foundation. Furthermore, DlP acknowledges partial support by NKFIH
Grant K134946. AB acknowledges support from Deutsche Forschungsgemeinschaft (DFG, German Research Foundation) -- TRR 352 -- Project-ID 470903074. AS's work was supported in part by the CARIPARO Foundation Grant under grant n.~68079.
\pagebreak

\begin{appendix}
\numberwithin{equation}{section}
\section{Further details on the convergence of the numerical algorithm}
\label{app:numerics}
In this section, we would like to comment more on the algorithm used to solve the coupled system of TBA and exact Bethe equations, that we briefly introduce in the main text. The system is solved via an iterative scheme, whose detailed presentation can be found in \cite{Brollo:2023rgp}. Unlike our previous work, the solution of the equations for odd $M$ requires a more sophisticated implementation, due to the polynomial growth~\eqref{eq:divergentY} of the $Y_0(\gamma)$ and $Y(\gamma)$ functions.
Let us denote the functions describing the asymptotic solution in~\eqref{eq:divergentY} as $A_0^c(\gamma)$ and $A^c(\gamma)$, where the apex keeps track of the dependence on $c$.
The convolution integrals can be regularized by splitting them into two terms,
\begin{equation}
\int d\gamma' \ln\left( 1+Y_0(\gamma')\right)s(\gamma'-\gamma) =
\int d\gamma' \ln\left( \frac{1+Y_0(\gamma')}{1+A_0^c(\gamma')}\right)s(\gamma'-\gamma)
+ \int d\gamma' \ln\left( 1+A_0^c(\gamma')\right)s(\gamma'-\gamma)\, .
\end{equation}
With this prescription, it is possible to choose a cut-off $\Lambda$ such that
\begin{equation}\label{cutoff}
\lim_{\gamma\to\Lambda} \ln\left( \frac{1+Y_0(\gamma)}{1+A_0^c(\gamma)}\right) = 1 + \mathit{err} \, ,
\end{equation}
where $\mathit{err}$ is exponentially small for sufficiently large values of $\Lambda$.
In this way, the first integrand on the right-hand side is negligible for $|\gamma|>\Lambda$, while the second one can be evaluated numerically without difficulty.
According to this prescription, we introduce the \emph{reduced} Y-functions $y_0(\gamma)$ and $y(\gamma)$ through $Y_0(\gamma)=y_0(\gamma)A_0^c(\gamma)$ and $Y(\gamma)=y(\gamma)A^c(\gamma)$.
Accordingly, equation~\eqref{eq:SimplifiedTBA_excited} can be rewritten as
\begin{equation}\label{cured new tba}
\begin{aligned}
    &\ln y_0(\gamma) = -L \tilde{E}(\gamma)
    +  \left(\ln{\left(\frac{1-A^cy}{1-A^c}\right)^4} *s\right)(\gamma)
    + \left(\ln{\left(1-A^c\right)^4} *s\right)(\gamma)
    - \ln A_0^c(\gamma), \\
    &\ln y(\gamma) =
    \left(\ln{\left(\frac{1+A_0^cy_0}{1+A_0^c}\right)} *s\right)(\gamma)
   + \left(\ln{(1+A_0^c)} *s\right)(\gamma)
   - \ln A^c(\gamma)\\
   &\qquad\qquad\quad + \sum_{j=1}^{M} \ln\left( S_{*}(-\gamma_{2j}-\gamma)S_{*}(\gamma_{2j}-\gamma)\right)\,,
\end{aligned}
\end{equation}
where $y_0(\gamma), y(\gamma) \to 1$ for $|\gamma|\gg c$. In these equations  only regularised convolutions appear, and they can be safely computed using FFT.

\begin{figure}[t]
    \centering
    \includegraphics[width=0.8\linewidth]{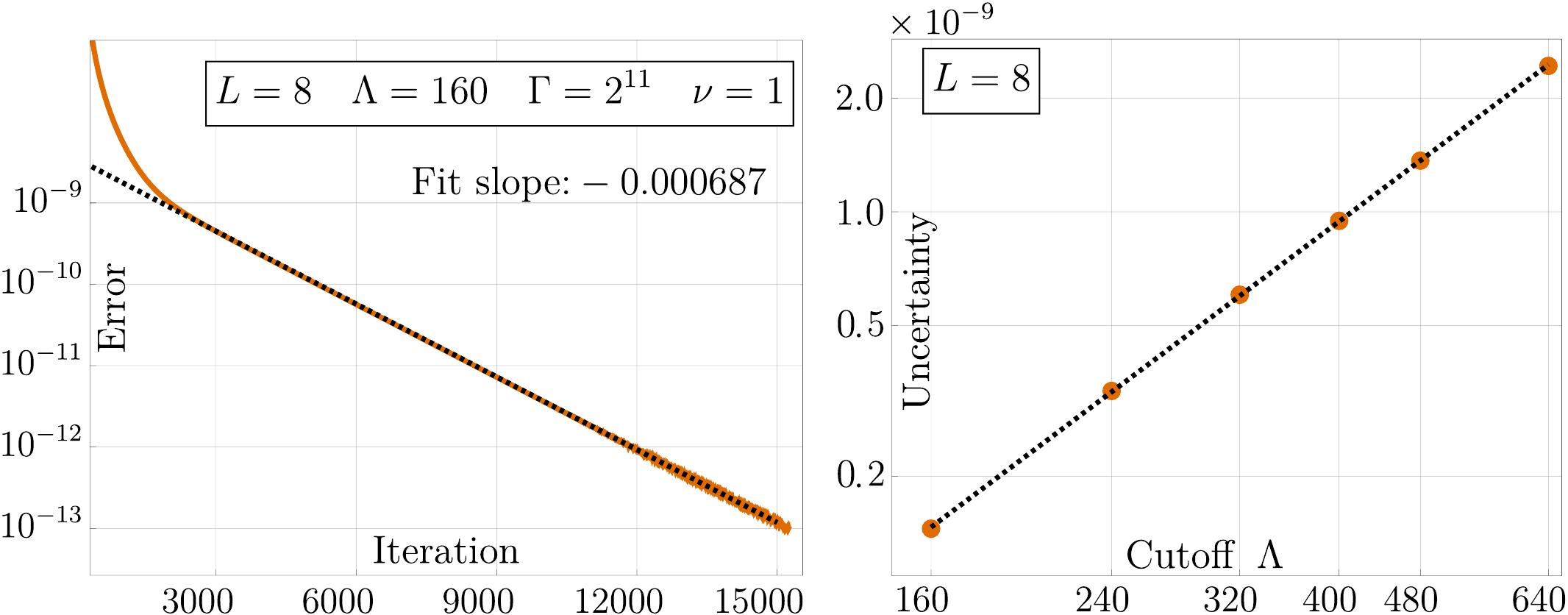}
    \caption{\textbf{Iterative scheme.}  On the left panel, the convergence of the iterative scheme is represented. On the $y$-axis, the difference between two consecutive values of the energy is shown, while on the $x$-axis the corresponding iteration number is reported. After an initial polynomial decay, the convergence becomes exponential, although slow. The slope of the linear fit in log scale is also reported on the plot. On the right panel, the dependence of the uncertainty with respect to the cutoff is shown. The plot is in log--log scale, and the polynomial dependence clearly appears.}
    \label{fig:iterations}
\end{figure}

The crucial limitation of the numerical implementation of the above scheme is that the correct value of $c$ is not known \emph{a priori}. An incorrect choice of this parameter would cause the error in~\eqref{cutoff} to be only power-law suppressed, $\mathit{err}=\mathcal{O}(\Lambda^{-1})$, thereby potentially introducing non-negligible contributions for finite values of the cut-off. We addressed this issue by introducing a scheme in which, at each iterative step, the evolved Y-functions are fitted in order to update the parameter $c$. In particular, we choose to fit $Y_0(\gamma)$. This procedure minimizes the error in the convolutions, at the cost of a more complicated and slower algorithm.
Actually, as shown in Fig.~\ref{fig:iterations}, after an initial polynomial convergence, the iteration becomes exponential. Nevertheless, it requires $\mathcal{O}(10^5)$ iterations, whereas the method described in~\cite{Brollo:2023rgp} converges in $\mathcal{O}(10)$ steps. This significant slow-down of the iterative procedure raises concerns about the numerical precision of the final result.

\begin{figure}[t]
    \centering
    \includegraphics[width=0.8\linewidth]{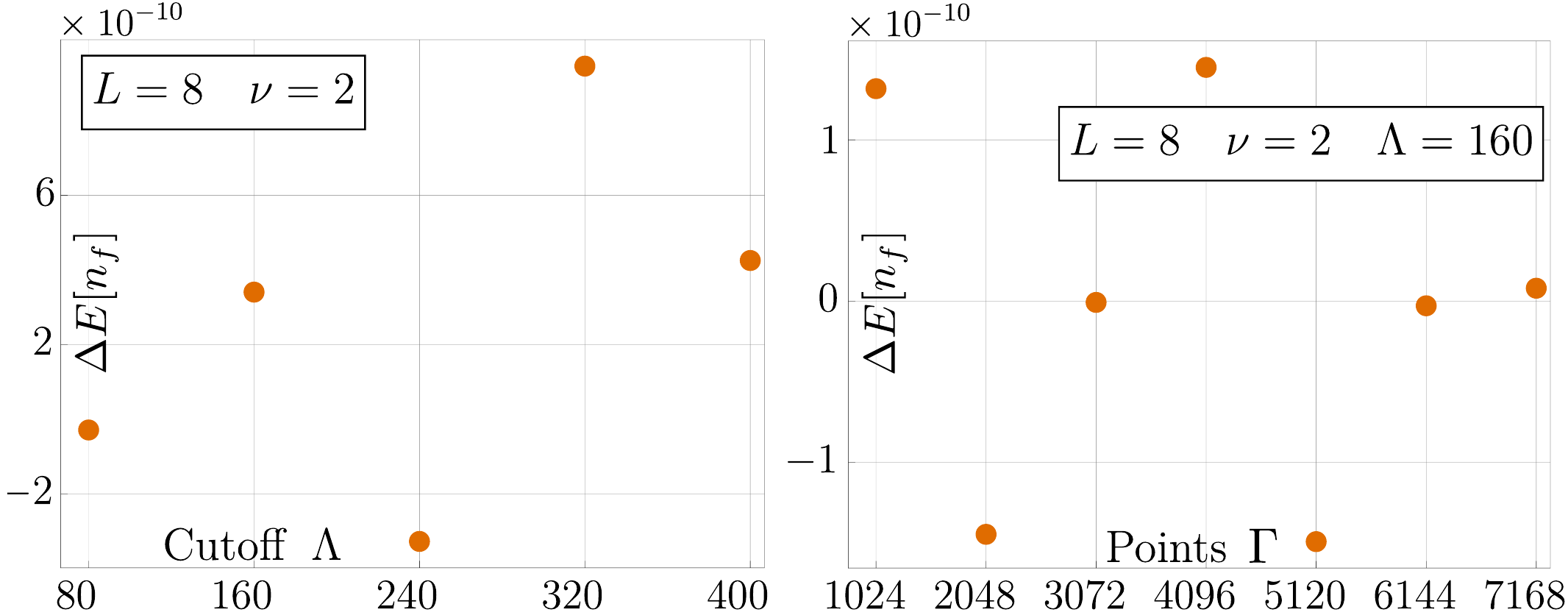}
    \caption{\textbf{Finite-size and discretization errors.} As discussed in the text, we plot $\Delta E[n_f]$ as a function of both the cutoff $\Lambda$ and the number of points in the discretization $\Gamma$. No structure appears, and the points seem to be mere fluctuations below the estimated uncertainty.}
    \label{fig:cutoff}
\end{figure}

Let us focus on the exponential regime of the convergence, and model it as $E[n] = E + a\,\mathrm{e}^{-b n}$, where $n$ labels the iteration.
An exponential fit of $E[n]$ itself is not a viable option, since it would not allow us to reliably associate uncertainties to the fitted parameters.
Denoting the final iteration by $n_f$, our best estimate for the energy is $E[n_f]$, and we wish to estimate its uncertainty, namely the deviation from the exact value, $E - E[n_f]$.
The iterative procedure is stopped when two consecutive values of the energy differ by less than $10^{-13}$.
However, due to the slow convergence, this condition does not guarantee that $E - E[n_f] < 10^{-13}$.
Indeed, defining $\mathit{err}[n] = E[n] - E[n-1] = a(1 - \mathrm{e}^{b})\,\mathrm{e}^{-b n}$,
we can extract an estimate of the uncertainty through the following procedure:
\begin{enumerate}
    \item We perform a linear fit of $\ln\!\big(\mathit{err}[n]\big)$ as a function of $n$, from which we extract the slope and intercept, $m = -b$ and $q = \ln\!\big(a(1 - \mathrm{e}^{b})\big)$.
    \item Under the assumption $E[n] = E + a\,\mathrm{e}^{-b n}$, the uncertainty on $E[n_f]$ is given by $E - E[n_f] = a\,\mathrm{e}^{-b n_f}$. 
    Expressed in terms of the fitting parameters, this becomes
    \begin{equation}\label{uncertainty}
        \text{Uncertainty}
        = \frac{\exp(m n_f + q)}{1 - \mathrm{e}^{-m}}
        = \frac{\mathit{err}[n_f]}{1 - \mathrm{e}^{-m}}
        = \frac{10^{-13}}{1 - \mathrm{e}^{-m}}.
    \end{equation}
\end{enumerate}
This procedure does not allow us to extract the exact value of the energy, since the convergence is not perfectly exponential.
Nevertheless, it provides a reliable estimate of the uncertainty associated with~$E[n_f]$.
A natural question is how the uncertainty depends on the choice of the cutoff $\Lambda$ and on the number of points $\Gamma$ used in the discretization. For the latter, we find negligible effects, whereas we observe a non-trivial dependence on the cutoff, as illustrated in figure~\ref{fig:cutoff}. Increasing the cutoff increases the number of iterations required for convergence and, as discussed in~\eqref{uncertainty}, leads to a larger uncertainty on~$E[n_f]$.

Finally, one must verify whether the value of $E[n_f]$ itself is affected by the choices of $\Lambda$ and $\Gamma$. The results of this analysis are presented in figure~\ref{fig:iterations}. Concerning the discretization, we fix $L = 8$, $\Lambda = 160$, and $\nu_j = \pm2$ (the behaviour being similar for other choices), and we study how $E[n_f]$ depends on $\Gamma$.
In particular, we sample $E[n_f](\Gamma)$ for equally spaced values $\Gamma_k = k~\Gamma_0$, and we consider the differences $\Delta E[n_f](k) = E[n_f]((k+1)~\Gamma_0) - E[n_f](k~\Gamma_0)$. This procedure removes any constant offset and allows us to directly probe the convergence pattern: the presence of exponential or polynomial trends would be immediately visible in the behaviour of $\Delta E[n_f](k)$.
As shown in figure~\ref{fig:cutoff}, no discernible structure is observed, and the fluctuations of $\Delta E[n_f](k)$ are fully consistent with the effective numerical precision discussed above. We therefore conclude that the error associated with the discretization is below the accuracy achievable by the iterative scheme.
A similar analysis is performed by fixing $L = 8$ and varying the cutoff $\Lambda$, while keeping the ratio $\Lambda/\Gamma$ constant, so that the discretization is unchanged. The results, shown in figure~\ref{fig:iterations}, again exhibit no systematic structure, indicating that the error induced by the cutoff is smaller than the effective precision reached by the iterative algorithm.

All things considered, we can safely conclude that the algorithm does not introduce significant systematic errors associated with either the discretization or the finite extent of the rapidity space. The dominant limitation to the accuracy of our results is instead related to the slow convergence of the iterative procedure. Nevertheless, the analysis presented above allows us to reliably estimate the residual uncertainty. In practice, we choose $\Lambda = 160$ and $\Gamma = 2048$, which ensure a final uncertainty on the energy of order $10^{-10}$.
As a final remark, we stress that a simpler scheme in which the asymptotic behaviour is factored out using an \emph{a priori} fixed value of the non-universal parameter $c$, without updating it during the iterations, would lead to a substantial loss of precision. In that case, we found systematic errors of order $\mathcal{O}(10^{-6})$, thereby preventing a reliable control of the numerical accuracy. This highlights the necessity of the adaptive procedure adopted in the present work.

\end{appendix}
\bibliographystyle{JHEP}
\bibliography{letter.bib}

\end{document}